\documentclass[letter]{aa} 

\usepackage{graphicx}
\usepackage{epstopdf}
\usepackage[varg]{txfonts}
\usepackage{hyperref}
\usepackage{siunitx}
\DeclareSIUnit{\solarmass}{M_\sun}
\DeclareSIUnit{\year}{yr}

\begin{document}

   \title{Planetary Nebulae with Ultra-Violet Imaging Telescope (UVIT): Far Ultra-violet halo around the Bow Tie nebula (NGC 40)\thanks{Based on data obtained with the Ultra-Violet Imaging Telescope (UVIT) on the ASTROSAT satellite.}}

   \author{N. Kameswara Rao \inst{1}\and F. Sutaria\inst{1}\and J. Murthy\inst{1}\and S. Krishna \inst{1}\and R. Mohan \inst{1}\and A. Ray\inst{2,3}}

   \institute{Indian Institute of Astrophysics, Koramangala II Block, Bangalore-560034, India\\
   \email{nkrao@iiap.res.in}
   \and
   Tata Institute of Fundamental Research, Colaba, Mumbai-400005, India
   \and
   Homi Bhabha Centre for Science Education (TIFR), Mumbai-400088, India}

   \date{Received October 27, 2017; accepted November 17, 2017}

 
  \abstract
   {\object{NGC 40} is a  planetary nebula with diffuse X-ray emission, suggesting an interaction of the 
   high speed wind from WC8 central star (CS) with the nebula. It shows strong \ion{C}{iv} 1550~\AA\ 
  emission that cannot be explained by thermal processes alone. We present here the first map of this
  nebula in \ion{C}{iv} emission, using broad band filters on the {\it UVIT}.}
   {To map the hot \ion{C}{iv} emitting gas and its correspondence with soft X-ray (0.3-8 keV)
    emitting regions, in order to study the shock interaction with the nebula and the ISM. This also illustrates the potential of
    {\it UVIT} for nebular studies.}
   {Morphological study of images of the nebula obtained at an angular resolution
  of about 1.3~\arcsec\ in  four UVIT filter bands that include \ion{C}{iv} 1550~\AA\ and \ion{C}{ii}] 2326~\AA\ lines and UV continuum. Comparisons with X-ray, optical, and IR images from literature.}
   {The \ion{C}{ii}] 2326~\AA\  images show the core of the nebula with two lobes on 
  either side of CS  similar to [\ion{N}{ii}].
     The \ion{C}{iv} emission in the core  shows similar morphology and extant
  as that of diffuse X-ray emission concentrated in nebular condensations.
           A surprising UVIT discovery is the presence of a large faint FUV halo in FUV Filter with $\lambda_{\rm eff}$ of 1608~\AA. 
  The UV halo is not present in 
   any other UV filter.  FUV halo is most likely due to UV fluorescence
   emission from the Lyman bands of H$_2$ molecules. Unlike the optical and IR halo, FUV halo
   trails predominantly towards south-east side of the nebular core, opposite to the
   CS's proper motion direction. }
   {Morphological similarity of \ion{C}{iv} 1550~\AA\ and X-ray emission in the
  core suggests that it results mostly from  interaction of strong CS wind with the nebula.   
  The FUV halo in \object{NGC 40} highlights the existence of H$_2$ molecules 
   extensively in the regions even beyond the optical and IR halos. Thus UV studies
 are important to estimate the amount of H$_2$, probably the  most dominant molecule
   and significant for mass loss studies.}


   \titlerunning{Ultra-violet halo around NGC40}
   \authorrunning{N.K.Rao et~al.}
   \maketitle
%
\section{Introduction}
           The Bow Tie nebula (\object{NGC 40}, PN G357.4-07.2) is a low excitation planetary 
nebula with a Wolf-Rayet central star (CS) of WC8 type that blows a strong high
 velocity stellar wind of \SI{1730}{\kilo\metre\per\second} \citep{fiebelman99} into a mildly expanding nebula. The star is hydrogen deficient, carbon rich and has T$_{\rm eff}$ of \SI{71}{kK} \citep{marcolino07} with high mass
 loss rate of about \num{e-5} to \SI{e-6}{\solarmass\per\year} \citep{bianchi92}.  It has been suggested that the hot CS is
   shielded by a carbon curtain which lowers the level of ionization in the nebula \citep{bianchi87}.

          \object{NGC 40} is a well studied nebula in all wavelengths. Morphological studies in 
 the optical and IR show that the nebula has a central core of about \SI{48}{\arcsec} diameter,
  which is described as 
 cylindrical or barrel-like in the IR and in low excitation lines, like [\ion{N}{ii}]
  \SI{6583}{\angstrom}, with
 two lobes on slightly south east and north west sides to the central star, the latter lobe
  being brighter. In high excitation lines like [\ion{O}{iii}] \SI{5007}{\angstrom} the central
 core is elliptical or almost circular with maximum emission within the  [\ion{N}{ii}]
  \SI{6583}{\angstrom} zone \citep{meaburn96,ramoslarios11}. The core is surrounded
 by two faint halos that are seen only in deep images in low excitation lines and in IR
  but not in high excitation lines like [\ion{O}{iii}] \SI{5007}{\angstrom}. The smooth inner halo
 might extend between \SIrange{48}{90}{\arcsec} in diameter where as the outer halo, filamentary and
 patchy,  might extend to a diameter of 4 arcminutes surrounding the star \citep[see][for a sketch]{meaburn96}. A prominent filament bright
 in H$\alpha$ $+$ [\ion{N}{ii}] exists on the north-east of the core. It has been shown
 by deep monochromatic images in H$\alpha$ $+$ [\ion{N}{ii}] and in IR using 
 unsharp masking techniques that the halo contains several almost circular rings of
  emission surrounding the central star and the nebular core with spacings of \SIrange{5}{7}{\arcsec}
 \citep{corradi04,ramoslarios11}, a remnant of the asymptotic giant
 branch (AGB) wind of earlier evolution. The IR
 imagery also shows that there are  emission spokes almost radial to the core present
 in the halo. The ISO
 and Spitzer spectra of the nebula show that it has \SIrange{90}{150}{\kelvin} cool dust, polycyclic 
 aromatic hydrocarbon(PAH) features apart from some low ionization ionic lines \citep{ramoslarios11}. \citet{martin02} have interpreted the bright filament in
 the north-east and other filaments and tongues as a result of interaction of the
 nebula as it moves in the interstellar medium (ISM).

           The discovery of diffuse soft X-ray emission from \object{NGC 40} confirmed the
 interaction of the fast stellar wind with slow moving AGB envelope \citep{montez05}.
 This emission arises from a hot bubble created by quasi-spherical fast wind shock and
  is distributed  in an annulus inside the nebular core (Fig. \ref{chandra}, left, blue patches) in patches, not completely filled in. 

           Several authors \citep{clegg83,pottasch03,zhang05} have analysed the optical and UV spectra (from International Ultraviolet Explorer [IUE]) and showed that the nebula is rich in carbon as well as in $^{19}$F
 strongly suggesting the mixing of nucleosynthetic products from AGB into the nebula.

\begin{figure}[h]
\begin{center}
\includegraphics[height= 8cm,width= 8cm]{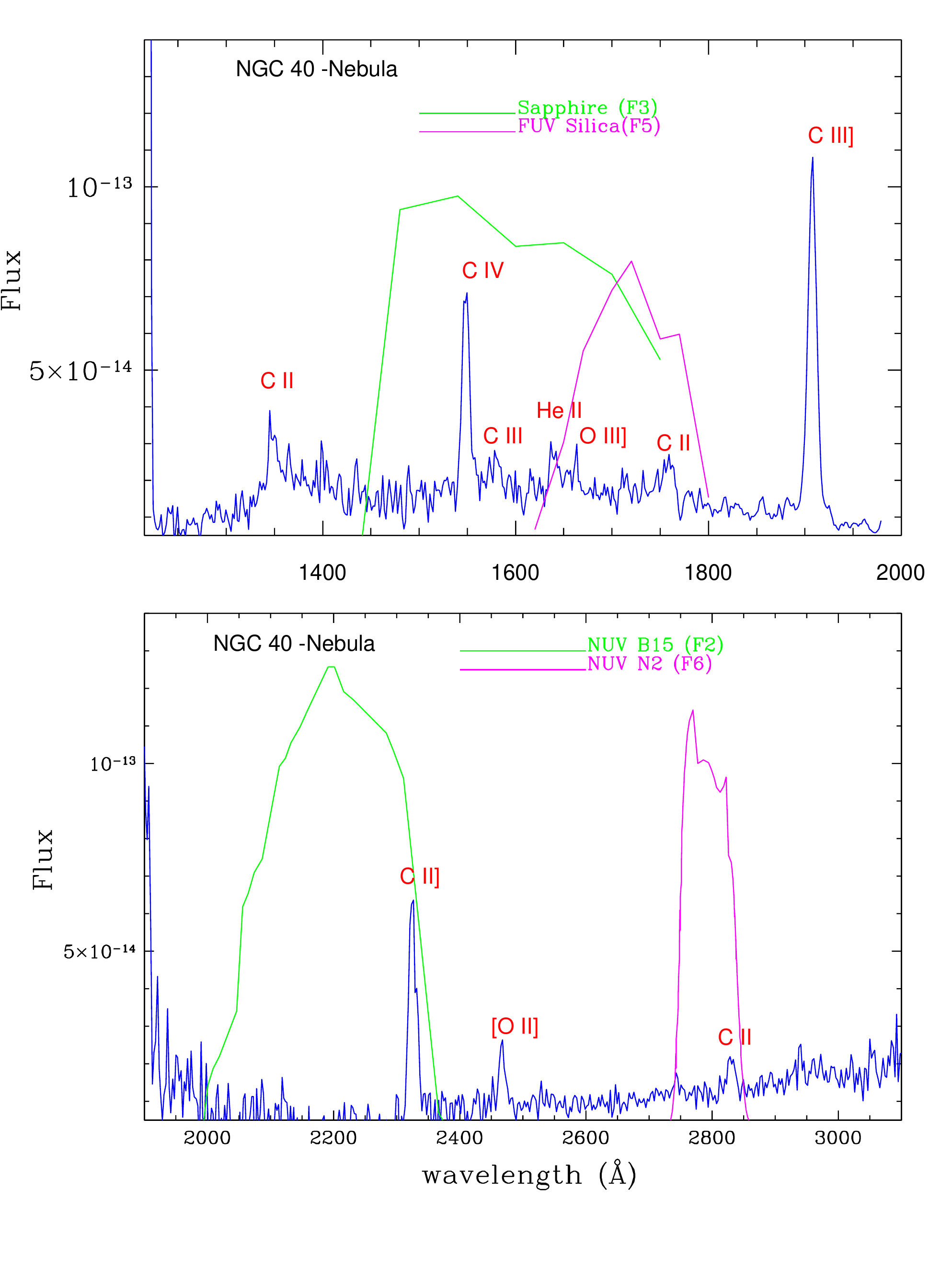}
\caption{IUE low resolution nebular spectrum of \object{NGC 40} (blue line) obtained
  at \SI{6}{\arcsec} north and \SI{13}{\arcsec} west of the central star. FUV is plotted on top and 
  NUV at bottom with UVIT filter effective areas (relative) shown. In the top panel green denotes F169M FUV Sapphire and magenta F172M FUV Silica, in bottom green is N219M NUV B15 and magenta N279N NUV N2.}
\label{iue}
\end{center}
\end{figure}

         \citet{clegg83}, from their 
 UV spectrum, point out that the \ion{C}{iv} \SI{1550}{\angstrom} emission is too strong in comparison with other low excitation nebulae and is unlikely to have been
  produced by thermal processes alone. They
  suggest it could be related to the strong stellar wind. This is one of the
  aspects, along with the desire to map the hot gas and its location with respect
  to X-ray emitting regions, that  prompted us to observe \object{NGC 40} with UVIT.  One of
  the far UV filters allows strong \ion{C}{iv} \SI{1550}{\angstrom}, \ion{He}{ii} \SI{1640}{\angstrom}) 
  and \ion{O}{III]} emission lines(\citet{fiebelman99}) and another filter admits the neighbouring continuum (Fig. \ref{iue}). \object{NGC 40} is one of the first planetary nebulae to be imaged with UVIT. We present our results including
  the surprising discovery of an FUV halo in this paper.

\section{ Observations}
  
           UVIT is one of the payloads on the Indian astronomical satellite ASTROSAT
 that was launched on 2015 September 28. It consists of twin telescopes of 38 cm 
 aperture  each. One of them is optimised for FUV and the other for NUV and optical:
 a dichroic mirror reflects the NUV and transmits the optical light. It operates
  in three channels FUV, NUV and optical simultaneously providing a 28 arc minute
 diameter field with a spatial resolution of about 1.3 arc sec in UV bands. The two UV 
 channels are provided with five filters and a low resolution transmission grating.
 The details of the instrument are given in \citet{Kumar2012} and in orbit performance is given in \citet{tandon17a}.
 
 The instrument was aimed at RA: 00$^h$13$^m$01.0$^s$, Dec: \ang{+72;31;19.1}. The present observations of 
 \object{NGC 40} have been obtained on 2016 December 9 in two FUV filters, F169M (Sapphire: 
  $\lambda_{\rm eff}$ of \SI{1608}{\angstrom}) and F172M (Silica: 
$\lambda_{\rm eff}$ of \SI{1717}{\angstrom})  as well as in two NUV filters, N219M (B15: 
  $\lambda_{\rm eff}$ of \SI{2196}{\angstrom}) and N279N (N2: $\lambda_{\rm eff}$ of \SI{2792}{\angstrom}). The effective exposure times that went into building the images  are
 \SI{1102}{\second} for F169M FUV Sapphire, \SI{1339}{\second} for F172M FUV Silica,  \SI{1294}{\second} for N219M NUV B15,
 and 1071s for N279N NUV N2. The stellar images in F169M filter show a PSF of 1.4 arcsec.

In its standard operating mode, UVIT will take images of the sky with a frame rate of \SI{29}{s^{-1}} which are stored on board and then sent to the Indian Space Science Data Centre (ISSDC) where the data are written into instrument specific Level 1 data files. \citet{murthy17} has written a set of procedures (JUDE) to read the Level 1 data, extract the photon events from each frame, correct for spacecraft motion (image registration) and add into an image. We used Astrometry.net \citep{Astrometry2010}  for an astrometric calibration and \citet{Rahna_UVIT_2017} for a photometric calibration. We have coadded the individual images and placed them all on a common reference frame and these were used for our further scientific analysis.

        Fig. \ref{iue} shows the IUE spectrum of \object{NGC 40} nebula with the transmission functions 
(arbitrary scale) of the filters we used. It is clear that F169M FUV Sapphire allows high excitation lines of
 \ion{C}{iv} \SI{1550}{\angstrom}, \ion{C}{III} \SI{1576}{\angstrom}, \ion{He}{II} \SI{1640}{\angstrom}, \ion{O}{III] \SI{1661}{\angstrom}} emission and F172M FUV Silica the continuum and weak \ion{C}{II} \SI{1761}{\angstrom}. Similarly the NUV 
 filter N219M  allows \ion{C}{ii}] \SI{2328}{\angstrom} emission and filter N279N has mostly the continuum
 although the weak \ion{C}{ii} \SI{2836}{\angstrom} might also contribute.

\begin{figure}[h]
\begin{center}
\includegraphics[height= 9cm, width= 9cm]{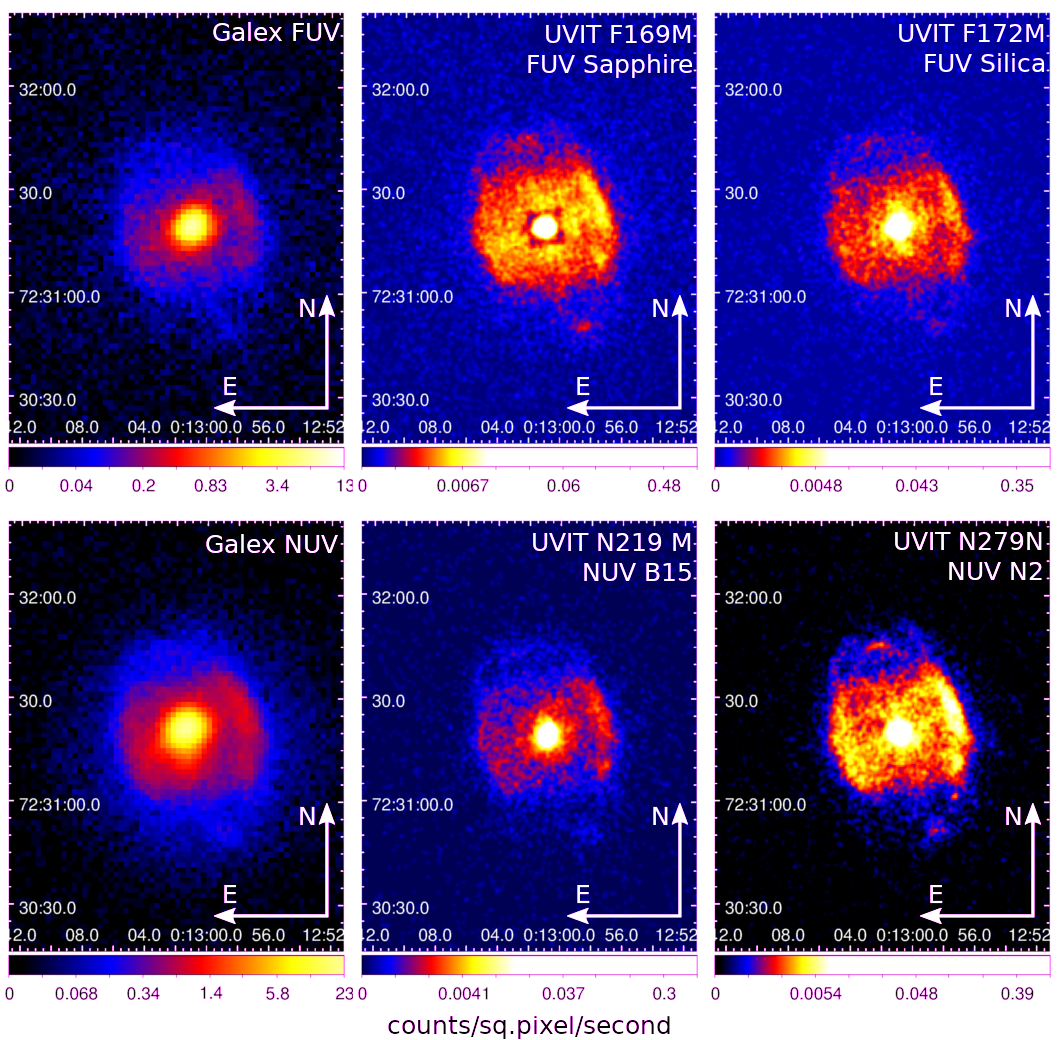}
\caption{ Comparison of Galex images of \object{NGC 40} with UVIT images. Note that the UVIT images provide better resolution and show structure in the nebula. The core of \object{NGC 40} in 
 F169M FUV Sapphire (top middle) 
 is compared with N219M NUV B15 that shows mainly \ion{C}{ii}] \SI{2326}{\angstrom} emission. \ion{C}{ii}] image is very similar to low excitation nebular line emissions
 seen in [\ion{S}{ii}] or [\ion{N}{ii}] \citep{meaburn96} where as F169M
  image shows several bright nebular condensations (also see Fig. \ref{chandra}).The axes are $\alpha$ and
  $\delta$}
\label{Galexcomp}
\end{center}
\end{figure}

\section{Results and Discussion} 


\begin{figure}[h]
\begin{center}
\includegraphics[height=4.5cm, width=9cm]{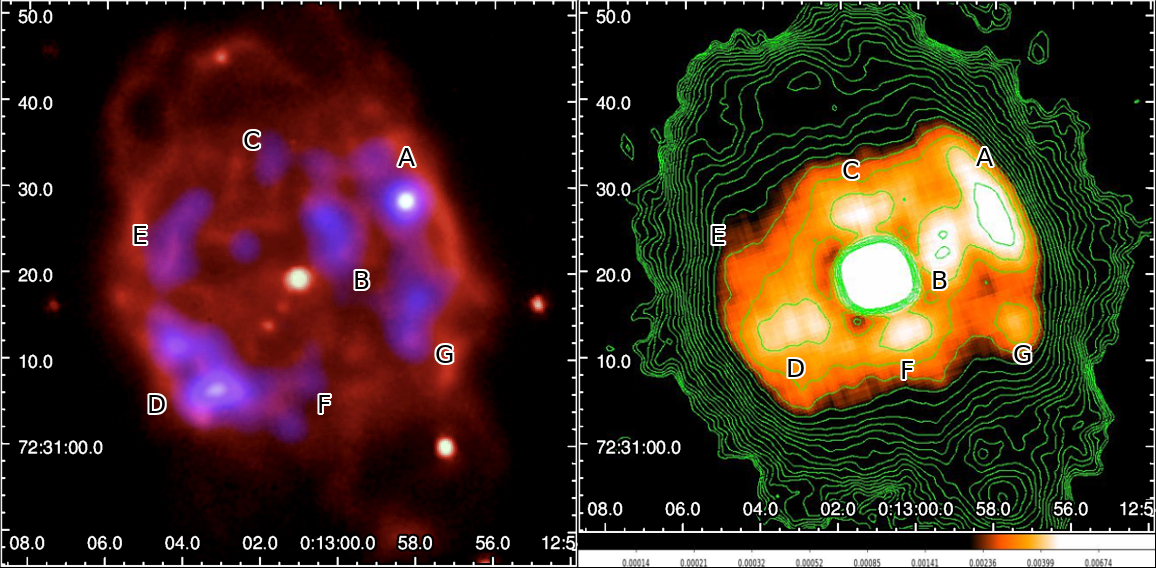}
\caption{X-ray image (left panel -blue emissions) superposed on H$\alpha$ image
 is compared with the contoured F169M FUV Sapphire image (right panel).
 Note the similarities of the bright condensations of emission in both images and the
over all elliptical shape.}
\label{chandra}
\end{center}
\end{figure}

\subsection{Morphological description}

               In Fig. \ref{Galexcomp} we present UVIT images in all observed filters along with Galex images in NUV 
  and FUV. It is evident that UVIT images provide highier spatial resolution  
  and show structure . The N219M  filter essentially displays the monochromatic image in
  \ion{C}{ii}] \SI{2326}{\angstrom} emission and F172M filter  shows nebula mostly in continuum  at
  \SI{1717}{\angstrom}.  The Galex FUV image shows an extension towards north-east along the axis of
  the nebula which is not present in UVIT FUV filters.
 
             \ion{C}{ii}] image (Fig. \ref{Galexcomp}, bottom center) looks very similar to the low
 excitation nebular
 line images (eg.[\ion{N}{ii}]) shown by \citet{meaburn96}. The barrel like
 structure is well seen with two bright lobes on either side of the barrel axis, the
  west one being brighter. 
  Along the barrel axis two condensations (ansae) are present on either side of the star at a
  separation of about \SI{32}{\arcsec}. The southern condensation is slightly brighter. 
  The extent of the image perpendicular to
 the barrel axis is about \SI{45}{\arcsec}. In contrast the FUV image, F172M, although
  similar to \ion{C}{ii}] image  is more filled-in. The lobes are not as prominent
 but the condensations on either side of the star are more prominent. However
 the FUV F169M
 image  shows a contrasting structure (Fig. \ref{Galexcomp}, top center) the bright emission region
 is more elliptical and there are four bright patches of nebulosity with distinct
 gaps. The southern condensations along the barrel axis is more conspicuous and a
  nebular loop connects that to the nebula. The size perpendicular to the barrel axis
  seem to be about \SI{54}{\arcsec}, larger than the \ion{C}{ii}] size. From the photoionization
 modelling \citep{monteiro11} it is expected that \ion{C}{iv}, \ion{He}{II},
 \ion{O}{III]} emission
 zones  to be smaller than \ion{C}{ii}] contrary to what is observed and confirms the
 statement by \citet{clegg83}.

\begin{figure}[h]
\begin{center}
\includegraphics[height=11.5cm, width=8.5cm]{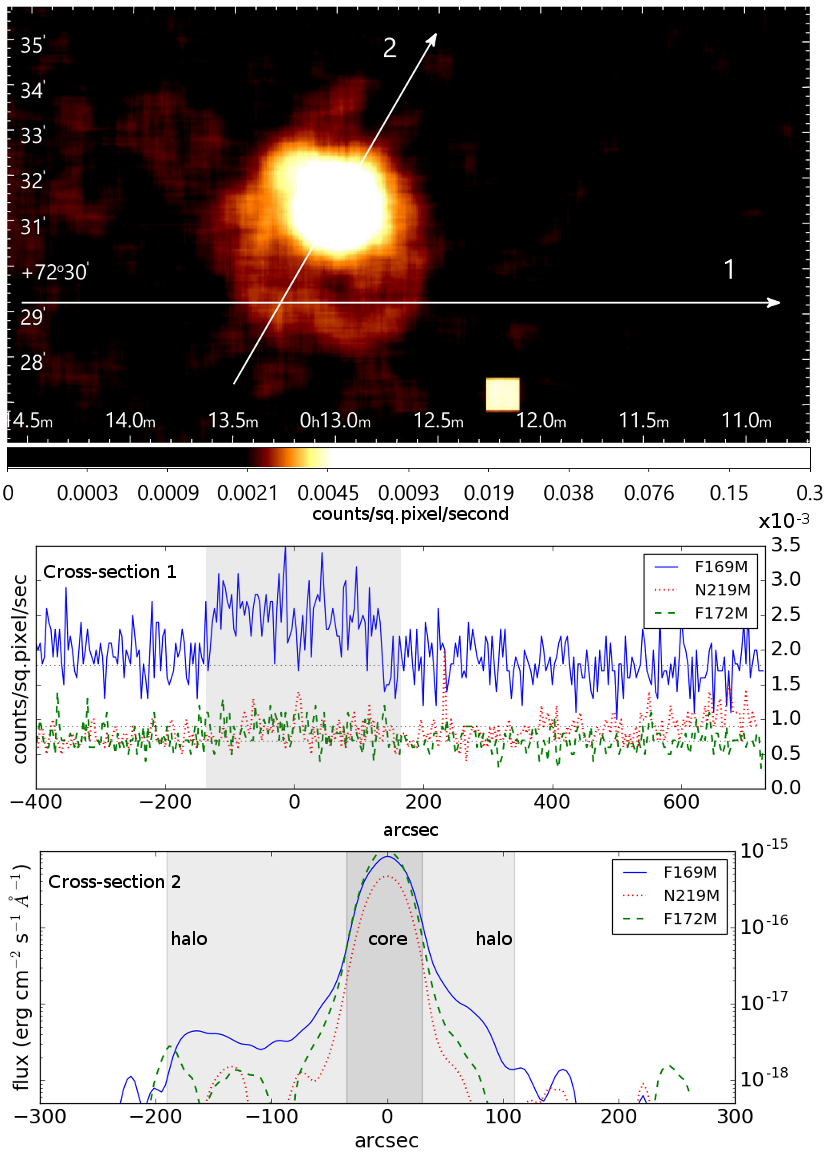}
\caption{ Figure shows cross-cuts in the images of the nebula in three filters F169M (blue), F172M (green) and N219M (red) made along the white lines shown in the slightly smoothed F169M image of the nebula (top). Cross-cuts of 8 pixel widths were obtained in all the images at the same locations (coordinates). The bottom cross-cut, which passes through the nebular core, is on a logarithmic scale and has been smoothed. Note the faint quasi-circular extended halo  around the core  of the
 nebula (on the south-east side). This region has been shown with a grey overlay in the plots. Despite having similar core fluxes in all three filters, only F169M shows the presence of the halo.}
\label{HaloCross}
\end{center}
\end{figure}

           \ion{C}{iv}, with its ionization potential of \SI{48}{eV}, might be the highest ionized
  species seen in the spectrum (\ion{He}{ii} lines are very weak). We compared the
  F169M core image with the soft X-ray image produced by \citet{montez05} in 
Fig. \ref{chandra} (\url{http://chandra.harvard.edu/photo/2005/n40/}). 
   We show contoured and enlarged F169M image of the nebular core 
  highlighting the bright regions. The similarity in the images is obvious. The emission
  in both X-rays and in high excitation lines exists  in patches and both are distributed in an 
  annulus. These emissions seem to follow X-ray emission and thus partly comes
  from shocked regions (the hot bubble). Modelling \ion{C}{iv} and X-ray emission together would provide strong constraints to 3D models. 
   It is of interest to note that the low resolution slit-less FUV grating spectra (Rao et~al. in preparation)
obtained with UVIT shows the \ion{C}{IV} \SI{1550}{\angstrom} nebular emission 
extending over the core.

\subsection{FUV Halo and ISM}

\begin{figure}[h]
\begin{center}
\includegraphics[height= 6cm, width= 9cm]{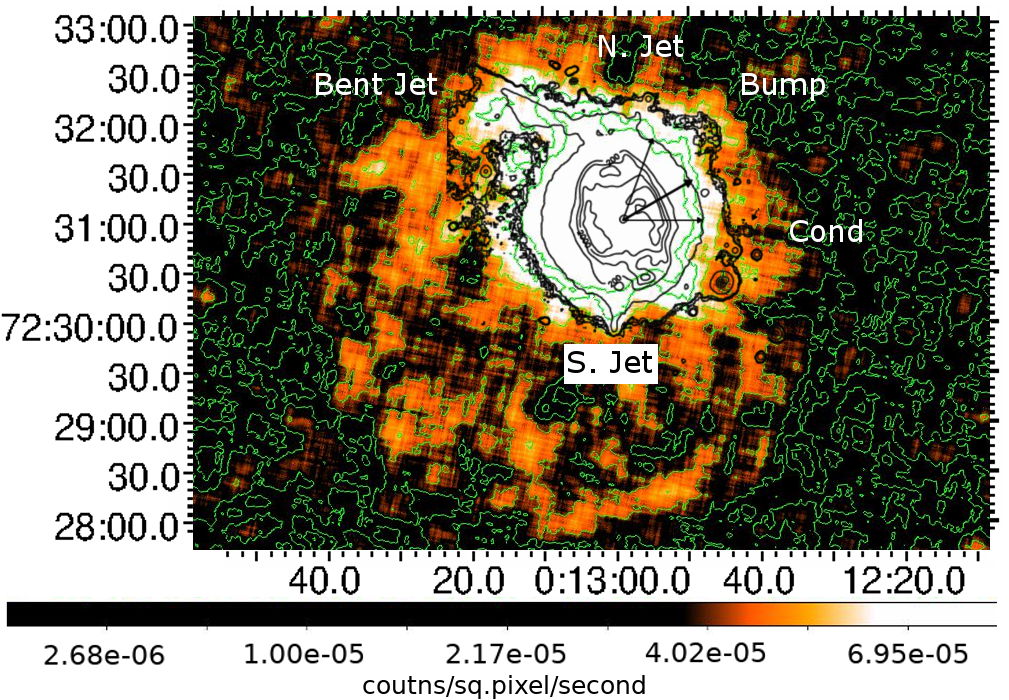}
\caption{  Sketch of contoured H$\alpha$ halo image illustrating interaction
 with ISM from Martin et al (2002) (dark contours) superposed on slightly smoothed FUV F169M image with green-contours showing
 the FUV halo. Note the similarities -the bent jet, N.jet and the bump. }
\label{FUVhalo}
\end{center}
\end{figure}

                 A comparison of the F169M FUV Sapphire image with the F172M FUV Silica shows faint
   halo around the core of the nebula in F169M FUV Sapphire mainly to south-east extending 
   to about 4 arc minutes (Fig. \ref{HaloCross}, top right) from the central star. This halo is not present in 
   any of the other 
   NUV filters nor in Galex FUV or NUV images. Thus the discovery of  a halo in F169M FUV Sapphire  is a total surprise although Galex FUV image shows
   an extension to north-east. Slightly processed image of the halo with  contours
   is shown in Fig. \ref{FUVhalo}.
   
   It is of some interest to see whether
 the absence of the halo in other UV filters is because of differences in 
 exposure times,  band width etc . 
  The band widths of F169M, F172M and N219M are  \SI{290}{\angstrom}, \SI{125}{\angstrom} and \SI{270}{\angstrom}
\citep{tandon17a} respectively. The 
 peak flux in the nebular core converted from counts/s to flux using \citep{tandon17b} in flight calibration seem to be similar in F169M and F172M
 and slightly less in N219M.  As Fig. \ref{HaloCross} (bottom plot) shows  the logarithamic flux plots of the nebula across the core
   in the three filters, it clearly illustrates the existence of a halo around the
   nebular core in F169M and its absence in the two filters F172M and N219M. The absence of halo in 
 other UV filters is not due to lack of 
 adequate exposure time or sensitivity.

            Two aspects are note worthy.(1) Unlike the optical and IR halos, which are more
  or less symmetrical around the core of the nebula \citep{meaburn96},  the FUV halo
  is mainly towards south-east side of the central star (Fig. \ref{FUVhalo}, bottom) and extends to about 
  4 arc minutes from the star.  FUV halo
  is also filamentary.  There are similarities with optical outer halo as well.
  The bent jet \citep{martin02} or the bands denoted as A and B by \citet{ramoslarios11}.
  (2) The axis of the
   FUV halo seems to point in the direction of proper motion of the star. Fig. \ref{FUVhalo}
  (black contours) from \citet{martin02} show the proper motion vector (black arrow).

             The source of emission of FUV halo can not be the \ion{C}{iv} \SI{1550}{\angstrom} line
  for the obvious reason that no high excitation line emission including [\ion{O}{iii}]
  \SI{5007}{\angstrom} is  seen in the halo \citep{meaburn96,corradi04} or outside
  of the nebular core. FUV Grating spectrum (Rao et~al. in preparation) also do not show any \ion{C}{IV} emission extending to the halo (however this might also be due to lack of sensitivity.
 
  The  dust scattering of UV photons from the \SI{71}{kK} central star
   might not be appropriate since dust (Mie) scattering would be over a broader wavelength
   range   thus the absence of halo 
  emission
   in the filter centered at \SI{1717}{\angstrom} can not be reconciled. Thus the only other
   possibility for  emission to be present only in
   the filter centered at \SI{1608}{\angstrom} seems to be, emission in  Lyman bands of H$_2$
   molecules. Warm H$_2$ from circum-nebular regions have already been detected around many PNs from FUSE spectra \citep{herald04,herald07,herald11, mccandliss07}.  Such H$_2$ emissions have also been seen in reflection nebula \object{IC 63} \citep{witt89}, the FUV bright tail of \object{Mira} \citep{martin07}, FUV rings around carbon
  star U Hya \citep{sanchez14} and ISM \citep{martin90}. Few H$_2$ rotational lines in 2.1-2.4 \si{\micro\metre} region
  have been detected in the western lobe of \object{NGC 40} \citep{hora99}. However 
 H$_2$ emission was not detected in the narrow band imagery. The spectra produced by
 UV fluorescent process in \object{IC 63}
 (also diffuse ISM) observed and modelled \citep{france05}
   shows strong emission blueward of \SI{1608}{\angstrom} peak and no emission longward of \SI{1650}{\angstrom} .
  Since the halo is of low excitation, impact excitation by hot electrons of H$_2$, i.e by shocks, seems unlikely. 
    
             Interaction of FUV halo with ISM seems to be strongly present. The halo
 is mostly on the trailing side of nebular head that is oriented towards proper motion
 direction,  unlike IR halo. The nebular head is brighter in that direction (Fig. \ref{FUVhalo})
  of proper motion similar to case B of \citet{wareing07}. The filamentary structure
  and the direction of the filaments
  suggests strong interaction of the stellar wind with ISM as described by \citet{martin02}. 

                The novel result this paper presents is the discovery of FUV halo
 excited by UV fluorescent H$_2$ molecules around \object{NGC 40}. Such extensive presence of H$_2$
 is not obvious from IR lines. UV halos are not uncommon in compact PNs. \object{NGC 2440} provides
  another example (in preparation). UV observations are important to estimate H$_2$ content
  in PNs.

\begin{acknowledgements}
  We are very thankful and appreciative of  the comments and suggestions made by   anonymous referee which helped us to improve the contents of the paper. 
  UVIT and ASTROSAT observatory development  took about two decades before launch. Several people from several agencies were involved in this effort. We would like to thank them
 all collectively. NKR and SK would like thank Department of Science and Technology for their support through grant
  SERB/F/2143/2016-17 `Aspects in Stellar and Galactic Evolution'. AR would like to thank the Deparment of Atomic Energy, Govt. of India for a Raja Ramanna Fellowship.

Some of the data presented in this paper were obtained from the Mikulski Archive for Space Telescopes (MAST). STScI is operated by the Association of Universities for Research in Astronomy, Inc.

\end{acknowledgements}


\end{document}